# Removing External Degrees of Freedom from Transition State Search Methods using Quaternions

*Marko Melander[1,2*], Kari Laasonen[1,2], Hannes Jónsson[3,4]*

1) COMP Centre of Excellence, Aalto University, FI-00076 Aalto, Finland

2) Department of Chemistry, Aalto University, FI-00076 Aalto, Finland

3) Department of Applied Physics, Aalto University, FI-00076 Aalto, Finland

4) Faculty of Physical Sciences, University of Iceland, 107 Reykjavík, Iceland

**ABSTRACT**

In finite systems, such as nanoparticles and gas-phase molecules, calculations of minimum energy paths (MEP) connecting initial and final states of transitions as well as searches for saddle points are complicated by the presence of external degrees of freedom, such as overall translation and rotation. A method based on quaternion algebra for removing the external degrees of freedom is described here and applied in calculations using two commonly used methods: the nudged elastic band (NEB) method for finding minimum energy paths and DIMER method for finding the minimum mode in minimum mode following searches of first order saddle points. With the quaternion approach, fewer images in the NEB are needed to represent MEPs accurately. In both NEB and DIMER calculations of finite systems, the number of iterations required to reach convergence is significantly reduced. The algorithms have been implemented in the Atomic Simulation Environment (ASE) open source software.

**Keywords:** Nudged Elastic Band, DIMER, quaternion, saddle point, transition.

## 1. INTRODUCTION

Chemical reactions, diffusion events and configurational changes of molecules are transitions from some initial arrangement of the atoms to another, from an initial state minimum on the energy surface to a final state minimum. The path of highest statistical weight between the initial and final states, the minimum energy path (MEP), identifies the mechanism while the maximum energy along the MEP, a first order saddle point on the energy surface, gives the activation energy of a thermally induced transition within the harmonic approximation to transition state theory (HTST), where the rate constant is estimated to be

$$k^{HTST} = k_0 \exp\left[\frac{-(E^{\ddagger} - E^{init})}{k_B T}\right] \qquad (1)$$

Here, $E^{\ddagger}$ is the energy of the saddle point and $E^{init}$ is the energy of the initial state. The pre-exponential factor, $k_0$, accounts for entropic effects and is computed from normal mode frequencies at the initial state minimum and at the saddle point, $k_B$ is the Boltzmann factor and $T$ the temperature of the heat bath. An MEP is defined as a path where the energy is at a minimum with respect to all degrees of freedom orthogonal to the path tangent. It is possible to find a saddle point without finding the whole MEP. Once a saddle point has been found, it is important, however, to verify that it represents the highest energy along the minimum energy path connecting the initial and final state minima. When the MEP and highest saddle point on the MEP have been found, both the mechanism and the HTST estimate of the rate of the transition can be determined.

The MEP is a special path on the $3N - 6$ dimensional potential energy surface (PES), $N$ being the number of atoms in a 3-dimensional system. At any point on the MEP, force acts only along the path while the energy is at a minimum in all other directions[1]. The distance traveled along the MEP is the logical choice of a reaction coordinate. The six degrees of freedom for translation and rotation do not affect the energy. Thus, the MEP should not include these external degrees of freedom.

NEB[1,2] and DIMER[3-6] methods are commonly used tools for finding saddle points. In most cases rotation and translation do not pose a problem for these methods since they are often applied to systems subject to periodic boundary conditions and/or some atoms are held at fixed positions to mimic semi-infinite crystal. These constraints effectively remove translation and rotation from the periodic system during saddle point searches. However, in molecular systems or nanoparticles, the external degrees of freedom can lead to problems. In both NEB and DIMER methods, force projections and energy minimizations are used to find the MEP and/or the saddle points. Finite systems without constraints can rotate and/or translate to avoid high energy regions, such as saddle point regions, and thereby slow down or even prevent convergence. In NEB calculations, the path can be lengthened arbitrarily this way, allowing the system to avoid regions where the energy increases. This can lead to convergence problems and eventually a poor description of the MEP.

Translation can be removed easily by fixing the center of mass for each configuration, see below. Rotation is more difficult as it cannot be removed using rigid body dynamics since the structure of the system changes during the calculation and the principal axes of rotation vary from one configuration to another. A simple approach[7] is to remove six degrees of freedom by freezing one atom, constraining the movement of a second atom to a plane and restricting the motion of a third atom to a line. However, this approach may lead to incorrect, or even unphysical MEPs if the interaction between the constrained atoms and the rest of the atoms is not strong enough. An approach by Bohner *et.al.*[8] for removing rotation and translation from an NEB calculation relies on modifying the energy minimization algorithm by eliminating the six smallest eigenvalues of the Jacobian matrix of NEB forces. This approach has good convergence properties but is computationally quite demanding since it requies finding the eigenvalues at each iteration. Rühle *et.al.*[9] have used an axis-angle presentation to carry out rigid-body rotations for removing the external degrees of freedom in NEB calculations. The method has mainly been applied in calculations of rigid molecules.

In the present article, we describe a simple method based on quaternion algebra for minimizing rotation of the system in each iteration of the NEB and DIMER methods. This removes the rotational freedom and leads to faster convergence to the MEP in the NEB method and to first order saddle points in the minimum mode following method (MMF) where the minimum mode is obtained by the Dimer method.

## 2. METHODS

### 2.1 The NEB method

The NEB method is commonly used in calculations of MEPs for reactions and diffusion events in or at the surfaces of solids. The goal is to identify the transition mechanism and to find a saddle point separating the known, fixed initial ($R_0$) and final ($R_N$) states. The path between the initial to final states is represented with a set of $N+1$ replicas of the system, referred to as 'images', $[R_0, R_1, R_2, ... R_N]$. The coordinates of the atoms in the $N-1$ intermediate images are modified with an optimization algorithm until the force perpendicular to the path has been zeroed. The NEB force, $F_i$, on an image is the sum of the perpendicular component of the real force obtained from some description of the atomic interactions, $F_i^\perp$, and a spring force which lies along the local tangent, $F_i^\parallel$ and holds the system images properly spaced along the MEP, preventing them from sliding down to the minima. By displacing the images in the direction of $F_i^\perp$ the perpendicular force on the path is zeroed

$$F_i = F_i^\parallel - F_i^\perp$$
$$F_i^\perp = \nabla E(R_i) - \nabla E(R_i) \cdot \hat{\tau}_i \, \hat{\tau}_i \qquad (2)$$
$$F_i^\parallel = k(|R_{i+1} - R_i| - |R_i - R_{i-1}|)\hat{\tau}_i$$

where a simple estimate of the local unit tangent is

$$\hat{\tau}_i = \frac{R_{i+1} - R_{i-1}}{|R_{i+1} - R_{i-1}|} \qquad (3)$$

A better estimate of the tangent can be obtained from the line segment connecting to the neighboring image with higher energy and taking a weighted average for images that are locally of either maximum or minimum energy[2].

If translation and/or rotation occurs during an NEB calculation, the force acting on the images does not represent only changes in the path that are needed to zero the perpendicular component, but also allow the image to move along the path towards lower energy. In practice, this can lead to slow convergence since the minimization of the energy will make images avoid high-energy regions, such as the region around the saddle point. Convergence may not be reached at all as the path can become arbitrarily long, making the spacing between images arbitrarily wide, and thereby leaving no images in the region near the saddle point. All images in the end would slide down to the local minima. While convergence of the NEB to some given tolerance may be reached before this complete sliding down of the images has occurred, resolution of the path will be reduced and extra iterations performed. Proper convergence, therefore, requires a scheme for removing both translation and rotation.

**2.2. The DIMER method**

The DIMER method is used to find the normal mode corresponding to a minimum eigenvalue of the Hessian locally without the evaluation of second derivatives of the energy. This minimum mode estimate can be used in the MMF algorithm for finding saddle points on the energy rim surrounding a given initial state minimum at $R_0$. No information about a possible final state is assumed. The DIMER uses only first derivatives of the potential energy for computing the minimum mode making it computationally efficient even for large systems and systems where analytical second derivatives are not available. The DIMER method for finding the minimum mode uses two replicas of the system, $R_1$ and $R_2$, separated by a small, fixed distance, $2\Delta R$. $R_1$ and $R_2$ are points in $3N$-dimensional space, where $N$ is the number of atoms in a 3-dimensional system, and are defined as

$$R_1 = R_0 + \Delta R \widehat{N} \text{ and } R_2 = R_0 - \Delta R \widehat{N} \quad (4)$$

$\widehat{N}$ is a unit vector defining the orientation of the dimer. A search for a saddle point consists of two steps, rotation of the dimer and then translation (not to be confused with overall translation and rotation of the system which are external degrees of freedom). First, a rotational force, $F_R$, is used for rotating the dimer to find the direction of the normal mode corresponding to the lowest eigenvalue. Once this direction has been found, a translational force, $F_T$, is applied to move the dimer along this mode, increasing the dimer energy in this direction while reducing the energy along all perpendicular directions. Repeated iterations of this algorithm lead to convergence onto a first order saddle point. The forces used are

$$F_R = (\nabla E_1 - \nabla E_2) + [(-\nabla E_1 + \nabla E_2) \cdot \widehat{N}]\widehat{N}$$
$$F_T = \nabla E_0 - 2(\nabla E_0 \cdot \widehat{N})\widehat{N} \quad (5)$$

with $\hat{N} = (R_1 - R_2)/2$. Details of the algorithms for performing the dimer rotation and translation can be found in references [3,5,6].

As defined above, $R_1$ and $R_2$ are points in $3N$-dimensional configuration space. They define the dimer orientation, the direction $\hat{N}$ and search direction for the dimer translation. However, only $3N - 6$ degrees of freedom are needed for defining the PES, which means that six redundant external degrees of freedom, namely overall translation and rotation, are included in $\hat{N}$. Both dimer rotation and translation forces, thus, have components of the external degrees of freedom during saddle point searches leading to incorrect description of the minimum normal mode vector and less than optimal approach to the saddle point. Below, a general scheme and detailed algorithm for removing these external degrees of freedom is described.

**2.3. General scheme for removing translational and rotational degrees of freedom**

The method for removing the overall translation and rotation closely follows that of Coutsias *et.al.*[10]. The goal is to find a rotation matrix $\mathcal{R}$ and a translation vector $g$, to minimize the distance between a set of target coordinates $y'_k$ and model coordinates $x'_k$. This means that the task can be cast as a minimization problem with an objective function given by a residual, $\mathcal{E}$, defined as

$$\mathcal{E} = \frac{1}{N} \sum_{k=1}^{N} |\mathcal{R} x'_k + g - y'_k|^2 \qquad (6)$$

For an NEB calculation, $y'_k$ gives the coordinates of atom $k$ of an image and $x'_k$ the coordinates of the atom in a neighboring image in the NEB path. The vector $g$ is used to make the center of mass of $x'_k$ and $y'_k$ coincide while the matrix $\mathcal{R}$ is used to rotate $x'_k$ so as to minimize rotational motion in going from $x'_k$ to $y'_k$.

First, the centers of mass, $\bar{x}$ and $\bar{y}$, of the two vector sets $\{x'_k\}$ and $\{y'_k\}$ are moved to the origin. The translation vector $\boldsymbol{g}$ is simply the difference

$$\boldsymbol{g} = \bar{y} - \bar{x} \tag{7}$$

The center of mass is placed at the origin by defining the relative vectors $\{x_k\}$ and $\{y_k\}$ as

$$x_k = x'_k - \bar{x} \; ; \quad y_k = y'_k - \bar{y} \tag{8}$$

The residual then becomes

$$\varepsilon = \frac{1}{N} \sum_{k=1}^{N} |\mathcal{R} x_k - y_k|^2 \tag{9}$$

At this point, quaternion algebra is introduced and a quaternion $q$ is defined as a 4-vector: $q = [q_0, q_1, q_2, q_3] = [q_0, \boldsymbol{q}]$. Ordinary Cartesian vectors, such as $x_k$ and $y_k$, can be written as pure quaternions with $q_0 = 0$:

$$x_k = [0, x_k] \tag{10}$$

and similarly for $y_k$. An important application of quaternions is their use as rotation operators. The rotation $\mathcal{R} x_k$ is then written as

$$\begin{aligned} x_k^R &= \mathcal{R} x_k \\ x_k^R &= [0, x_k^R] \\ x_k^R &= \hat{Q} x_k \hat{Q}^{-1} = [q_0, \boldsymbol{q}][0, x_k][q_0, -\boldsymbol{q}] \end{aligned} \tag{11}$$

where $\hat{Q}$ is a unit quaternion. From the above, we see that either $\mathcal{R}$ or $\hat{Q}$ can be used to describe rotation. The rotation matrix $\mathcal{R}$ can be written in terms of components of $\hat{Q}$ as

$$\mathcal{R} = \begin{pmatrix} (q_0^2 + q_1^2 - q_2^2 - q_3^2)/2 & q_1 q_2 - q_0 q_3 & q_1 q_3 + q_0 q_2 \\ q_1 q_2 - q_0 q_3 & (q_0^2 - q_1^2 + q_2^2 - q_3^2)/2 & q_2 q_3 - q_0 q_1 \\ q_1 q_3 + q_0 q_2 & q_2 q_3 - q_0 q_1 & (q_0^2 - q_1^2 - q_2^2 + q_3^2)/2 \end{pmatrix} \quad (12)$$

Now, the task is to find the $\hat{Q}$ that minimizes the residual written in terms of quaternions, $\mathcal{E}q$, derived from Eq. (9)

$$\begin{aligned}
N\mathcal{E}q &= \sum_{k=1}^{N} (\hat{Q}^{-1}[\mathbf{0}, x_k]\hat{Q} - y_k)(\hat{Q}^{-1}[\mathbf{0}, x_k]\hat{Q} - y_k)^{-1} \\
&= \sum_{k=1}^{N} (|x_k|^2 - |y_k|^2) - 2L^T \mathcal{F} L
\end{aligned} \quad (13)$$

where $L = (l_0, l_1, l_2, l_3)$ is a 4-vector presentation of a quaternion while $\mathcal{F}$ is a matrix with elements of a correlation matrix $\mathcal{C}$:

$$\mathcal{C} = \sum_{k=1}^{N} x_k y_k^T \rightarrow \mathcal{C}_{ij} = \sum_{k=1}^{N} x_{ik} y_{jk} \; ; i, j = 1, 2, 3 \quad (14)$$

$$\mathcal{F} = \begin{pmatrix} \mathcal{C}_{11} + \mathcal{C}_{22} + \mathcal{C}_{33} & \mathcal{C}_{23} - \mathcal{C}_{32} & \mathcal{C}_{31} - \mathcal{C}_{13} & \mathcal{C}_{12} - \mathcal{C}_{21} \\ \mathcal{C}_{23} - \mathcal{C}_{32} & \mathcal{C}_{11} - \mathcal{C}_{22} - \mathcal{C}_{33} & \mathcal{C}_{12} + \mathcal{C}_{21} & \mathcal{C}_{31} + \mathcal{C}_{13} \\ \mathcal{C}_{31} - \mathcal{C}_{13} & \mathcal{C}_{12} + \mathcal{C}_{21} & -\mathcal{C}_{11} + \mathcal{C}_{22} - \mathcal{C}_{33} & \mathcal{C}_{23} + \mathcal{C}_{32} \\ \mathcal{C}_{12} - \mathcal{C}_{21} & \mathcal{C}_{31} + \mathcal{C}_{13} & \mathcal{C}_{23} + \mathcal{C}_{32} & -\mathcal{C}_{11} - \mathcal{C}_{22} + \mathcal{C}_{33} \end{pmatrix} \quad (15)$$

From Eq.13 it can be seen that the residual, $\mathcal{E}q$, is minimized when $L^T \mathcal{F} L$ is at an extremum. $L^T \mathcal{F} L$ has the form of a Rayleigh quotient when $L^T L = 1$ which is the condition for a unit quaternion. The extremum of the Rayleigh quotient equals its largest eigenvalue. Thus, the minimum of Eq. 13 is found by solving the eigenvalue problem

$$\mathcal{F} L = \lambda L \quad (16)$$

The eigenvector corresponding to the largest eigenvalue is the unit quaternion, $\hat{Q}$, of Eq. 13. The corresponding rotation matrix can be formed from the component of $\hat{Q}$ as in Eq. 12.

### 2.4. Flow charts of the algorithms of NEB-TR and DIMER-TR

Below, we provide flow charts of the algorithm described above for removing translational and rotational degrees in NEB and DIMER calculations. The "TR" suffix is added to designate the modified methods so as to differentiate them from the original methods. Both the NEB-TR and the DIMER-TR methods have been implemented in the NEB and DIMER modules of the Atomic Simulation Environment (ASE)[11], which is a Python based interface to several quantum chemistry, density functional theory and empirical potential descriptions of atomic interactions.

*2.4.1. NEB-TR*

The goal is to minimize rotation between adjacent points along an NEB path between the fixed initial ($R_0$) and final ($R_N$.) states. The path is represented by a discrete set of images of the system, $[R_0, R_1, R_2, \ldots R_N]$.

1. Shift the center of mass of $R_0$ to the origin of the coordinate system. The shifted coordinates are given by $\{y_k\}$.

2. Shift the center of mass of $R_1$ to the origin. The shifted coordinates are given by $\{x_k\}$.

3. Compute matrices $\mathcal{C}$ and $\mathcal{F}$ from equations 14 and 15.

4. Solve the eigenvalue problem given by equation 16. The eigenvector corresponding to the largest eigenvalue defines $\hat{Q}$ in equation 13.

5. Using the elements of $\hat{Q}$, form the rotation matrix $\mathcal{R}$ in equation 12.

6. Rotate $x_k$ using $\mathcal{R}$ to obtain $\{x_k^R\}$ and $R_1^R$.

7. Move the center of mass of $R_0$ back to its original position ($\bar{y}$).

8. Replace $R_0$ with $R_1^R$ and repeat steps 2-7 for the remaining $M-1$ movable images along the path.

9. Compute the NEB forces using the rotated coordinates.

This scheme is applied to the initial path and to each intermediate path generated in the iterative NEB calculation.

*2.4.2. DIMER-TR*

In DIMER-TR, external translational and rotational degrees of freedom are removed from dimer translation and rotation steps by modifying the coordinates and the normal mode of equation 5.

1. Read the starting geometry $R_0$ and choose an initial (random) orientation $\widehat{N}$. Compute $R_1$ and $R_2$ in equation 4.

2. Shift the center of mass of $R_0$ to the origin of the coordinate system. The atomic coordinates with respect to this origin are $\{y_k\}$.

3. Shift the center of mass of $R_1$ to the origin. The atomic coordinates of $R_1$ then become $\{x_k\}$.

4. Compute matrices $\mathcal{C}$ and $\mathcal{F}$ from equations 14 and 15.

5. Solve the eigenvalue problem of equations 16. The eigenvector corresponding to the largest eigenvalue defines $\widehat{Q}$ in equation 13.

6. Using the elements of $\widehat{Q}$, form the rotation matrix $\mathcal{R}$ in equation 12.

7. Rotate $x_k$ using $\mathcal{R}$ to obtain $\{x_k^R\}$ and $R_1^R$.

8. Compute a new normal mode direction $\widehat{N}^R = (R_1^R - R_0)/\Delta R$.

9. Move the center of mass of $R_0$ back to its original position ($\bar{y}$). Compute new values of $R_1$ and $R_2$ using $\widehat{N}^R$ in equation 4.

10. Search for the optimal dimer orientation as detailed in references 3, 5, and 6. Carry out steps 1-9 for each new orientation.

11. After convergence to the minimum mode has been reached, translate the dimer according to algorithms detailed in references 3, 5, and 6 to obtain a new value of $\boldsymbol{R}_0$. Remove translations and rotations each time the dimer is displaced. Go to step 1.

## 3. RESULTS

Below, the NEB-TR is applied to a simple test problem, the reorganization of a Lennard-Jones tetramer ($LJ_4$), as well as the dissociation of a CO molecule on a $Fe_{13}$ nanoparticle. The DIMER-TR is applied to a gold island on a Pt nanoparticle. The performance is in both cases compared with calculations including external translation and rotation.

### 3.1. NEB-TR for Reorganization of $LJ_4$

To test the NEB-TR method, we first consider the reorganization of a tetramer cluster where the interaction between atoms is described by the Lennard-Jones potential

$$V_{LJ} = 4\epsilon \sum_{i<j} \left[ \left(\frac{\sigma}{r_{ij}}\right)^{12} - \left(\frac{\sigma}{r_{ij}}\right)^{6} \right] \quad (17)$$

Here, $r_{ij}$ is the distance between two particles and $\epsilon = \sigma = 1$, in the present case. Owing to the simplicity of the potential and small system size, both energy and forces can be computed quickly and accurately making the $LJ_4$ reorganization a convenient test case. The MEPs were calculated using 22 images and optimized using the FIRE algorithm[12]. Several computations using both regular NEB and NEB-TR with different force tolerance for defining convergence were performed. The results are summarized in Table 1.

**Table 1.** Results of calculations of minimum energy paths for rearrangements of a tetramer of atoms interacting with Lennard-Jones potential using both the NEB and NEB-TR methods with various convergence criteria for the atomic force, $f_{max}$ (in units of $\epsilon/\sigma$) and using up to 10 000 iterations. The number of iterations needed to reach convergence, the estimate obtained of the saddle point energy, $E^{\ddagger}$ (in units of $\epsilon$), and the path length (in units of $\sigma$) as defined in Eq. 18 are listed.

| | NEB-TR | | | NEB | | |
|---|---|---|---|---|---|---|
| $f_{max}[\epsilon/\sigma]$ | $E^{\ddagger}$ [$\epsilon$] | Iterations | Path length [$\sigma$] | $E^{\ddagger}$ [$\epsilon$] | Iterations | Path length [$\sigma$] |
| 1.0 | 1.516 | 52 | 5.633 | 2.766 | 79 | 7.55 |
| 0.1 | 0.931 | 68 | 5.628 | 0.935 | 321 | 20.984 |
| 0.01 | 0.926 | 88 | 5.622 | 0.926 | 895 | 10.719 |
| 0.001 | 0.926 | 421 | 5.622 | 0.926 | 2397 | 10.034 |
| 0.0001 | 0.926 | 773 | 5.622 | --- | --- | --- |

The initial and final states as well as the saddle points found for two different mechanisms of the configurational change of the LJ$_4$ cluster are shown in Figure 1. At the lower energy saddle point ($E^{\ddagger} = 0.926\ \epsilon$), an atom passes over a bridge site and the cluster takes the shape of a rhombus. The higher energy saddle point ($E^{\ddagger} = 2.77\epsilon$) is a triangle with one atom passing through the center of the triangle.

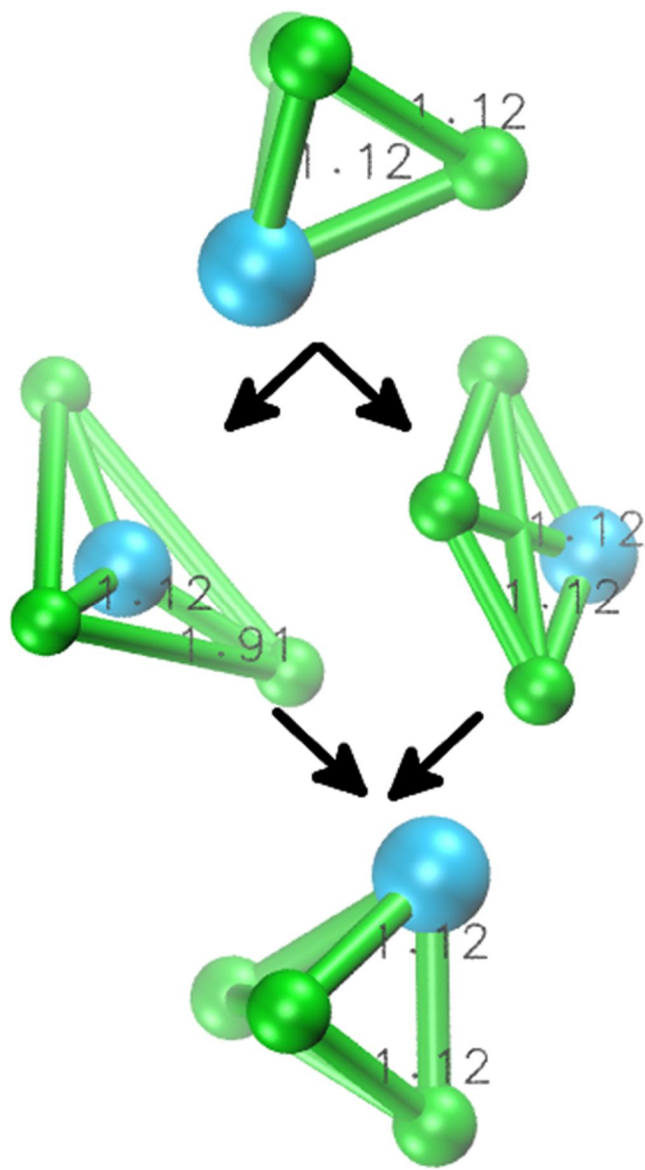

**Figure 1**. The initial (top) and final (bottom) arrangements of atoms in a tetramer cluster where the atoms interact according to the Lennard-Jones potential. The images in the middle show saddle point configurations: a triangular one with energy $E^\ddagger = 2.77\,\epsilon$ (left) and a rhombus with energy $E^\ddagger = 0.926\,\epsilon$ (right). The blue sphere indicates the atom that moves most. The numbers give bond lengths in units of $\sigma$.

From Table 1, it can be seen that the NEB-TR requires significantly fewer iterations for convergence than the regular NEB. When the tolerance on the maximum atomic force is reduced below $10^{-3}$ $\epsilon/\sigma$, the regular NEB does not converge. Furthermore, NEB-TR in all the calculations converges to the path corresponding to the lower energy saddle point, the rhombohedron. The regular NEB converges to the MEP with triangular, high-energy saddle point when the tolerance is large. In general, however, the NEB converges on the MEP that is nearest to the initial path.

Another clear distinction between the NEB and NEB-TR methods is the difference in the length of the path obtained, defined as

$$l_{path} = \sum_{i=0}^{M-1} \sqrt{(R_{i+1} - R_i)^2} \qquad (18)$$

Path lengths given in Table 1 show that NEB-TR always finds shorter MEPs than the regular NEB. The regular NEB gives a path that is longer than the MEPs because translation and rotation are included. The effect of translation and rotation can also be seen in Figure 2 where several NEB images have the same energy as the initial and final structures; the images with equal energy correspond to the translated or rotated images that have managed to slide down as a result of the force minimization. In NEB-TR, all neighboring images have different energy. The path obtained with the NEB-TR is clearly closer to the MEP than the path obtained with the regular NEB.

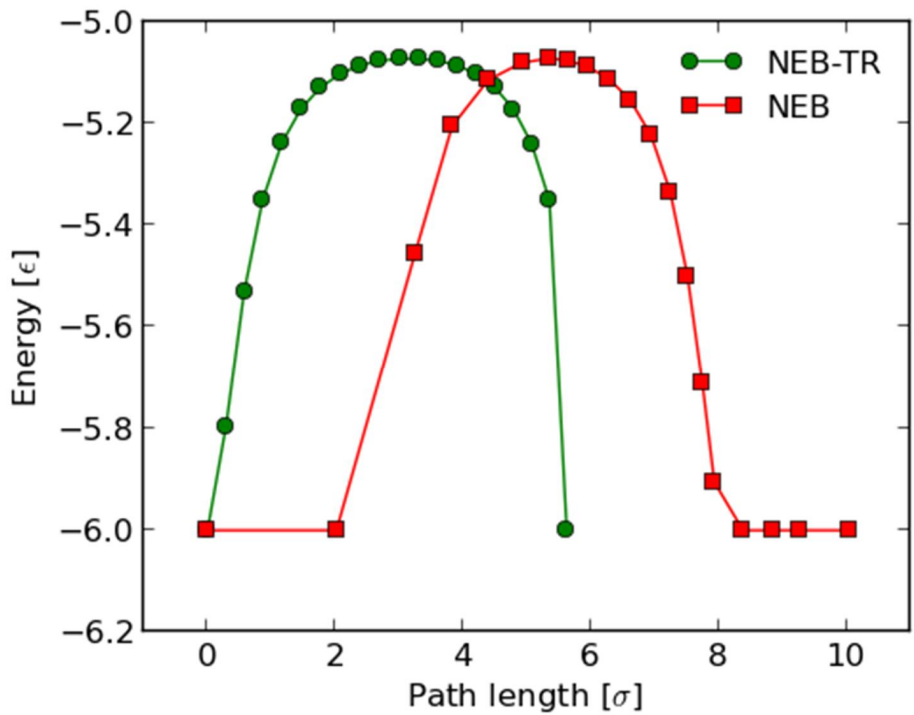

**Figure 2.** Paths obtained using the regular NEB and NEB-TR methods for the rearrangement of a tetramer cluster where the atomic interactions are described by the Lennard-Jones potential. The convergence criterion was a tolerance of $f_{max} = 10^{-3}\ \epsilon/\sigma$ in the atomic forces.

### 3.2. NEB-TR for CO Dissociation on Fe13

In an earlier study[13], CO dissociation on an icosahedral Fe$_{13}$ nanocluster was observed to be a particularly difficult path to calculate with NEB; convergence was poor and additional images around the saddle point needed to be added, increasing the total number of images to 15. Here, we compare NEB and NEB-TR for this reaction using DFT within the projector augmented-wave method as implemented in the GPAW software.[14,15] The paths were optimized using the FIRE algorithm[12]. The DFT calculations were carried out using a real-space grid with 0.18 Å grid-spacing. The PBE[16] energy functional was used. Spin-polarization was included in all the calculations and the magnetic momentum was allowed to relax during the SCF cycles. The SCF convergence threshold for the wave functions was $10^{-6}$. The cluster was modeled in a non-periodic cubic simulation cell with edge length of 24 Å, giving at least 10 Å distance between atom centers and box edges in all directions. The energy and force on the highest energy image in the NEB after 50 iterations using either 8 or 15 images to represent the path are given in Figure 3. The 8 image paths are shown in Figure 4.

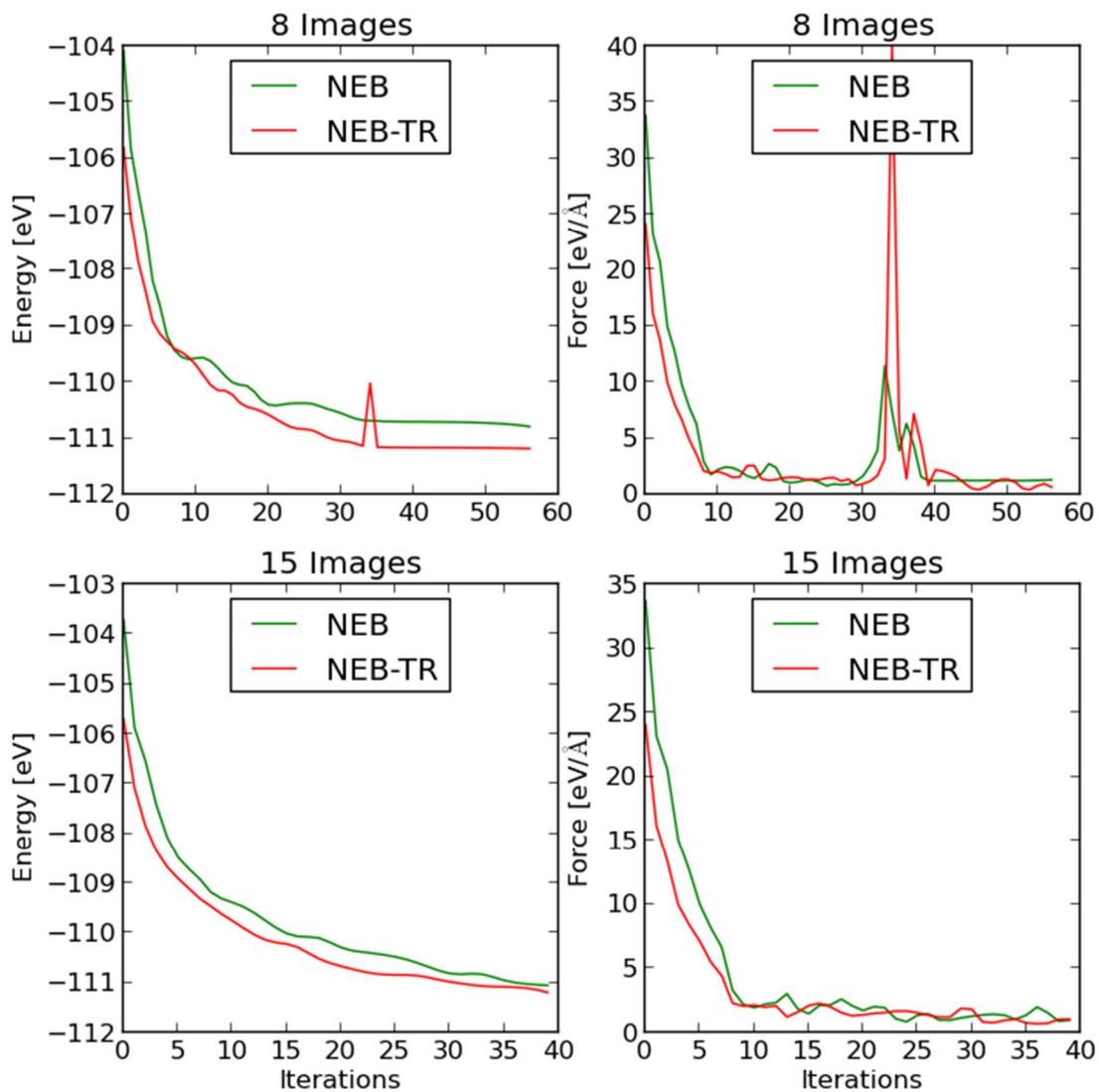

**Figure 3**. Energy and atomic force as a function of iteration number in NEB and NEB-TR calculations of CO dissociation on a $Fe_{13}$ cluster using 8 images (upper panel) or 15 images (lower panel).

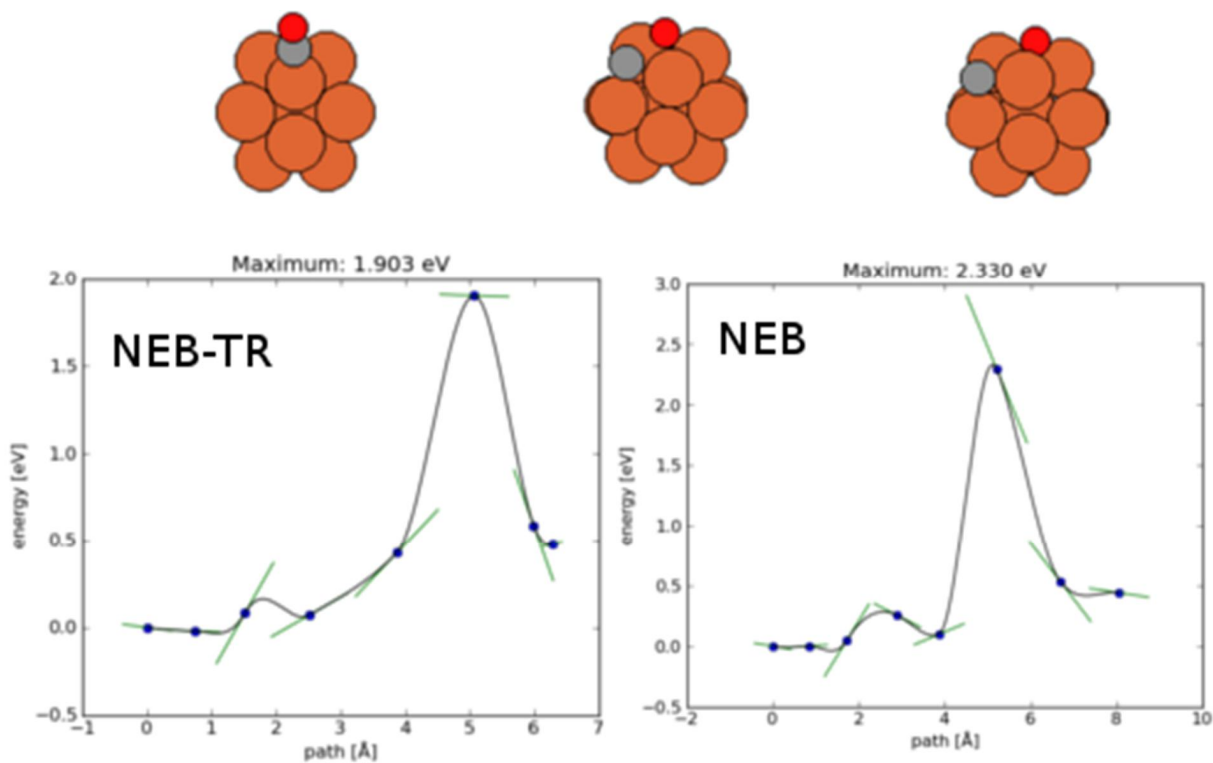

**Figure 4**. Paths obtained for CO dissociation on a $Fe_{13}$ cluster from NEB and NEB-TR methods after 50 iterations. The green line segments indicate the force calculated along the tangent to the path. The initial, saddle point and final state structures obtained from the NEB-TR calculation are shown above.

The NEB-TR calculations all reach lower energy than the regular NEB calculations, as can be seen from Fig. 3. Also, the atomic forces are smaller for NEB-TR especially during the first 8 iteration steps which is due to a better initial straight line interpolation path between the initial and final state configurations with the NEB-TR method. Using 8 images, the NEB-TR path converges to energy of -111.18 eV while the energy of the NEB path is -110.75 eV. With 15 images, both NEB methods approach the value of -111.20 eV and give a CO dissociation energy barrier of 1.93 eV. Figure 4 shows that NEB-TR with 8 images has found a saddle point (flat tangent) with an energy of 1.90 eV in 50 iterations. However, the normal NEB struggles to find the saddle point and the highest energy image keeps drifting around with an energy of around 2.3 eV. The NEB-TR method also gives a more even spacing of images than the regular NEB method, as can be seen in Figure 4. The length of the path, as defined in Eq. 18, with both 8 and 15 images is shorter for NEB-TR (6.5Å and 8.1 Å) than for the regular NEB (8Å and 11.5Å). These results illustrate how the path becomes artificially long in the regular NEB calculations because of the additional flexibility provided by the external degrees of freedom (see equation 2-3). An accurate estimate of the saddle point energy can be obtained with fewer images when the NEB-TR method is used.

### 3.3. DIMER-TR for rearrangements of a Au trimer on $Pt_{55}$

The performance of the DIMER-TR method was tested by calculating saddle points of a $Au_3$ island on a cuboctahedral $Pt_{55}$ nanocluster using the empirical EMT[17] potential function provided in ASE. A total of 50 DIMER calculations with a maximum of 1000 translation steps in each one were performed starting from initial configurations generated with small random displacements of the gold atoms from their minimum energy configuration. The convergence criterion for the total force was 0.05 eV/Å. The initial state and some of the saddle points found are depicted in Figure 5. The number of iterations needed using NEB and NEB-TR methods to find the various saddle points are shown in Figure 6.

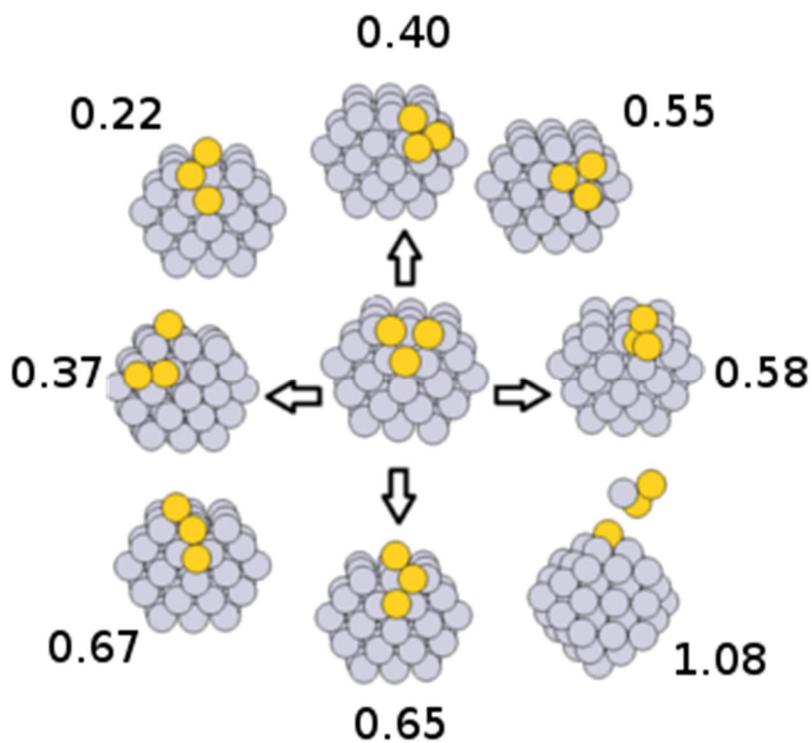

**Figure 5**. The initial configuration (center) and several different saddle point configurations found with the DIMER-TR method for the $Pt_{55}$-$Au_3$ island system. The energy of the saddle points with respect to the initial state minimum ranges from 0.22 eV to 1.08 eV, as indicated by the labels.

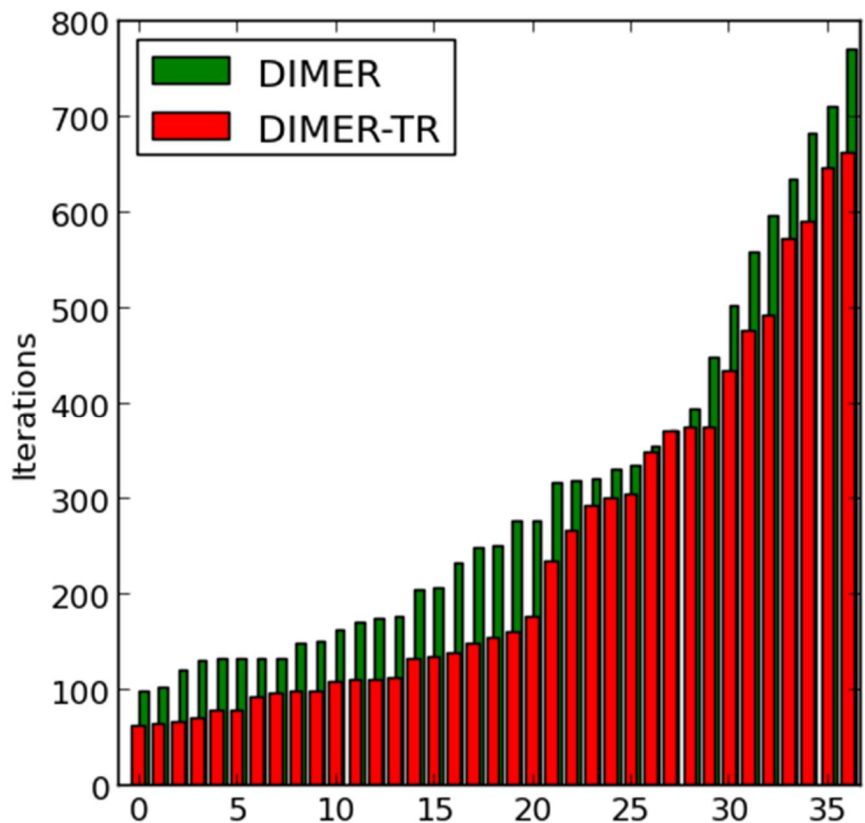

**Figure 6**. Comparison of the number of iterations needed for convergence in 50 saddle point searches for the $Pt_{55}$-$Au_3$ island system using the DIMER (including rotation and translation) and DIMER-TR methods. On average, the removal of rotation and translation reduced the number of iterations by 30%.

The results show that DIMER-TR consistently requires fewer iterations (up to 30%) than the uncorrected DIMER method to converge on a saddle point. Both methods converge to saddle points with roughly the same energy and structure. The DIMER-TR method, however, seems to find slightly more low-energy saddle points. However, using the same atomic displacements as an initial guess, both methods usually converge to the same saddle point with DIMER-TR requiring fewer iterations for convergence. There is no apparent correlation between the number of iterations required for convergence and the energy of the saddle point.

## 4. SUMMARY AND DISCUSSION

Overall translation and rotation of a finite system are external degrees of freedom that can hamper convergence and reduce the efficiency of MEP calculations and saddle point searches. We describe here an efficient method for removing the external degrees of freedom for two commonly used methods: NEB and DIMER. Translations are removed by keeping the center of mass stationary during the calculations. Rotations are removed by minimizing the distance between adjacent configurations of the atoms with a method based on quaternion algebra. Detailed algorithms are provided for the modified NEB-TR and DIMER-TR methods. Comparison of NEB-TR and regular NEB where translation and rotation are included shows that NEB-TR calculations need fewer images, converge faster and present the MEPs more accurately. In extreme cases, convergence is not reached without the removal of the external degrees of freedom. The DIMER-TR calculations consistently require about 30 % fewer iterations for convergence on a saddle point than a DIMER calculation that includes translation and rotation. We expect the NEB-TR and DIMER-TR methods to prove particularly useful in calculations of paths and saddle points in systems representing nanocatalysts and gas-phase chemical reactions. The algorithms described here have been made available in the ASE software and through that can be used in DFT calculations with various software packages, as well as in calculations using a range of empirical potential functions.

The NEB method is widely used to estimate MEPs for transitions in solids and the surfaces of solids[18] but its applicability to finite systems has been hampered by the problems associated with overall rotation of the system. The NEB-TR method presented here can, however, easily be applied to transitions involving gas phase molecules and nanoclusters. The results presented here show that not only does the NEB-TR method converge faster, but it also converges on a path that is closer to the MEP, i.e. the path for which the energy of the system is at a minimum with respect to all degrees of freedom perpendicular to the path. It is important to emphasize that when more than one MEP connects given initial and final state minima, as is the case for the rearrangement of the atoms in the Lennard-Jones tetramer (see Fig. 1) the NEB method tends to converge on the path closest to the initial path. Initial paths for the NEB method are typically generated by a linear interpolation between the endpoint configurations. But, a better initial path can be generated by fitting interpolation of changes in pairwise distances between atoms in the initial and final states[19]. More generally, a sampling of possible paths needs to be carried out to ensure that the optimal path has been found.

Typically the NEB converges to a discrete approximation of an MEP. There are, however, cases where the zeroing of the force perpendicular to the path does not necessarily bring the NEB to an MEP[20,21]. A zero gradient path is not necessarily an MEP[22]. Since the goal of an NEB calculation is typically to find the highest saddle point and obtain an estimate of the rate of a transition within the HTST approximation, the Hessian matrix at the point of maximum energy along the path is subsequently evaluated and diagonalized to evaluate the pre-exponential factor in the HTST expression. One and only one of the eigenvalues of the Hessian should be negative at a first order saddle point. If two eigenvalues are negative, then further relaxation along the unstable mode perpendicular to the path is required to converge on a first order saddle point. This is an important test, along with the requirement that all the atomic forces should vanish at the saddle point. Similar tests involving a Hessian evaluated at each of the images can be carried to ensure that the path found by the NEB coincides with an MEP, but in typical applications it is sufficient to verify that the highest energy point along the path is a first order saddle point.

The DIMER method is used in combination with the MMF method in simulations of long time scale dynamics of systems where the fast vibrational motion is eliminated and the evolution of the system is represented only by a sequence of transitions from one state to another [18,23]. The DIMER-TR method will make it easier to apply this approach to, for example, long time scale simulations of structural transitions in isolated nanoclusters. The DIMER-TR method could also be used in combination with various other methods, for example the newly developed basin constrained $\kappa$-dimer method[24] which reduces the tendency of the MMF method to find saddle points that are not connected directly to the initial state minimum. The DIMER-TR method can also be used in the context of global optimization where saddle point searches are used to move from one local minimum of an objective function to another[25,26].

ACKNOWLEDGEMENTS

Funding from the Academy of Finland through Project No. 140115, Centre of Excellence Program (Project No. 251748), and FiDiPro program (Grant No. 263294), as well as the Icelandic Research Fund are acknowledged. CSC IT Center of Science provided computing resources trough their Grand Challenge project, gc6260.

*Corresponding author: marko.melander@aalto.fi, TEL: +358 40 524 2869

REFERENCES

1. Henkelman, G; Uberuaga, B.P.; Jónsson, H. *J. Chem. Phys.* **2000**, *113*, 9901.

2. Henkelman, G.; Jónsson H. *J. Chem. Phys.* **2000**, *113*, 9978.

3. Henkelman, G.; Jónsson H. *J. Chem. Phys.* **1999**, *111*, 7010.

4. Olsen, R.A.; Kroes, G.J.; Henkelman, G.; Arnaldsson, A.; Jónsson H. *J. Chem. Phys.* **2004**, *121*, 9776.

5. Heyden, A.; Bell, A.T.; Keil, F. J. *J. Chem. Phys.* **2005**, *123*, 224101.

6. Kästner, J.; Sherwood, P. *J. Chem. Phys.* **2008**, *128*, 014106.

7. H. Jónsson, G. Mills, K.W. Jacobsen, Nudged elastic band method for finding minimum energy paths of transitions, In Classical and Quantum Dynamics in Condensed Matter Simulations B.J. Berne, G. Ciccotti, D.F. Coker, Ed.; World Scientific, Singapore, 1998; pp. 397.

8. Bohner, M.U.; Meisner, J.; Kästner, J. *J. Chem. Theory Comput.* **2013**, *9*, 3498-3504.

9. Rühle, V.; Kusumaatmaja, H.; Chakrabarti, D.; Wales, D. J. *J. Chem. Theory Comput.* **2013**, *9*, 4026-4034.

10. Coutsias, A.A.; Seok, C.; Dill, K.A. *J. Comp. Chem.* **2004**, *25*, 1849-1857.

11. Bahn, S.R.; Jacobsen, K.W. *Comput. Sci. Eng.* **2002**, *4*, 56-66.

12. Bitzek, E.; Koskinen, P.; Gähler, F.; Moseler, M.; Gumbsch, P. *Phys. Rev. Lett.* **2006**, *97*, 170201.

13. Melander, M.; Latsa, V.; Laasonen, K. *J. Chem. Phys.* **2013**, *139*, 164320.

14. Enkovaara, J.; Rostgaard, C.; Mortensen, J.J.; Chen, J.; Dulak, M.; Ferrighi, L.; Gavenholt, J.; Glinsvad, C.; Haikola, V.; Hansen, H.A.; Kristoffersen, H.H.; Kuisma, M.; Larsen, A.H.; Lehtovaara, L.; Ljunsberg, M.; Lopez-Acevedo, O.; Moses, P.G.; Ojanen, J.; Olsen, T. *J. Phys.: Condens. Matter,* **2010**, *22*, 253202.

15. Mortensen, J.J.; Hansen, L.B.; Jacobsen, K.W. *Phys. Rev. B* **2005**, *71*, 135109.

16. Perdew, J.P.; Burke, K.; Ernzerhof, M. *Phys. Rev. Lett.* **1996**, *77*, 3865.

17. Jacobsen, K.W.; Stoltze, P.; Nørskov, J.K. *Surf. Sci.* **1996**, *366*, 394.

18. H. Jónsson, *Proc. Natl. Acad. Sci. U.S.A.* **2011**, 108, 944.

19. S. Smidstrup, A. Pedersen, K. Stokbro and H. Jónsson, *J. Chem. Phys.* **2014**, 140, 214106.

20. D. Sheppard and G. Henkelman, *J. Comput. Chem.* **2011**, 32, 1769.

21. W. Quack and J. M. Bofil, *J. Comput. Chem.* **2010**, 31, 2526.

22. K. Fukui, *Acc. Chem. Res.* **1981**, 14, 363.

23. S.T. Chill, M. Welborn, R. Terrell, L. Zhang, J-C. Berthet, A. Pedersen, H. Jónsson and G. Henkelman, *Modelling and Simulation in Materials Science and Engineering* **2014**, 22, 055002.

24. Xiao, P.; Wu, Q.; Henkelman, G. *J. Chem. Phys.,* **2014**, 141,164111.

25. A. Pedersen, J-C. Berthet and H. Jónsson, *Lecture Notes in Comp. Sci.* **2012**, 7134, 34.

26. M. Plasencia, A. Pedersen, A. Arnaldsson, J-C. Berthet and H. Jónsson, *Computers and Geosciences* **2014**, 65, 110.